\documentclass[nofootinbib,prb,amsmath,amssymb,floatfix,preprint,prb]{revtex4-2}
\usepackage{graphicx}
\usepackage{dcolumn}
\usepackage{bm}
\usepackage{siunitx}
\usepackage{mathtools}
\usepackage{yhmath}
\usepackage{subcaption}
\usepackage[hidelinks]{hyperref}
\usepackage{todonotes}
\graphicspath{ {images/} }

\usepackage{comment}

\renewcommand{\L}{\mathcal{L}}

\newcommand{\Z}{\mathbb{Z}}

\newcommand{\p}{\partial}
\newcommand{\dx}{\, \mathrm{d}}

\renewcommand{\v}{\mathbf{v}}
\renewcommand{\Re}{\mathrm{Re}}

\newcommand{\ie}{\emph{i.e.}}

\newcommand{\ds}{\displaystyle}
\newcommand{\iu}{\mathrm{i}\mkern1mu}

\renewcommand{\Im}{\mathrm{Im}}

\newcommand{\bc}{\mathbf{c}}

\newcommand{\cA}{A}
\newcommand{\ca}{a}

\newcommand{\bJ}{\mathbf{J}}

\newcommand{\bS}{\mathbf{S}}
\newcommand{\bk}{\mathbf{k}}

\newcommand{\bx}{\mathbf{x}}
\newcommand{\bn}{\mathbf{n}}
\newcommand{\by}{\mathbf{y}}
\newcommand{\bz}{\mathbf{z}}

\newcommand{\bq}{\mathbf{q}}

\newcommand{\vin}{v^\textrm{in}}
\newcommand{\sv}{v}
\newcommand{\svin}{\sv^\textrm{in}}
\newcommand{\svsc}{\sv^\textrm{sc}}
\newcommand{\vsc}{v^\textrm{sc}}
\newcommand{\ein}{e^\textrm{in}}
\newcommand{\esc}{e^\textrm{sc}}
\newcommand{\Jin}{\bJ^\textrm{in}}
\newcommand{\Jsc}{\bJ^\textrm{sc}}
\newcommand{\Pin}{P^\textrm{in}}
\newcommand{\Psc}{P^\textrm{sc}}
\newcommand{\Pop}{P^\textrm{act}}
\newcommand{\Pext}{P}

\renewcommand{\overline}[1]{#1^*}

\makeatletter
\DeclareRobustCommand\vdots{%
	\mathpalette\@vdots{}%
}
\newcommand*{\@vdots}[2]{%
	\sbox0{$#1\cdotp\cdotp\cdotp\m@th$}%
	\sbox2{$#1.\m@th$}%
	\vbox{%
		\dimen@=\wd0 %
		\advance\dimen@ -3\ht2 %
		\kern.5\dimen@
		\dimen@=\wd2 %
		\advance\dimen@ -\ht2 %
		\dimen2=\wd0 %
		\advance\dimen2 -\dimen@
		\vbox to \dimen2{%
			\offinterlineskip
			\copy2 \vfill\copy2 \vfill\copy2 %
		}%
	}%
}
\DeclareRobustCommand\ddots{%
	\mathinner{%
		\mathpalette\@ddots{}%
		\mkern\thinmuskip
	}%
}
\newcommand*{\@ddots}[2]{%
	\sbox0{$#1\cdotp\cdotp\cdotp\m@th$}%
	\sbox2{$#1.\m@th$}%
	\vbox{%
		\dimen@=\wd0 %
		\advance\dimen@ -3\ht2 %
		\kern.5\dimen@
		\dimen@=\wd2 %
		\advance\dimen@ -\ht2 %
		\dimen2=\wd0 %
		\advance\dimen2 -\dimen@
		\vbox to \dimen2{%
			\offinterlineskip
			\hbox{$#1\mathpunct{.}\m@th$}%
			\vfill
			\hbox{$#1\mathpunct{\kern\wd2}\mathpunct{.}\m@th$}%
			\vfill
			\hbox{$#1\mathpunct{\kern\wd2}\mathpunct{\kern\wd2}\mathpunct{.}\m@th$}%
		}%
	}%
}
\makeatother

\begin{document}

\title{Optical theorem and generalized energy conservation for scattering of time-modulated waves}

\author{Erik Orvehed Hiltunen}
\affiliation{Department of Mathematics, University of Oslo, Norway} 
\email{erikhilt@math.uio.no}

\author{John C. Schotland}
\affiliation{Department of Mathematics and Department of Physics, Yale University, New Haven, CT USA} 
\email{john.schotland@yale.edu}

\date{\today}

\begin{abstract}
We introduce and study a generalized energy conservation relation for scattering of time-modulated waves, where conventional energy conservation does not hold. Based on this relation, we derive an optical theorem and compute the active power describing the power input or output due to scattering. Notably, the same system may be subject to energy gain, energy loss, or energy conservation depending on the frequency harmonics present in the wave field. Moreover, we show how the optical theorem derived herein may be used for image reconstruction based on measurements of the power. In particular, such measurements do not require information about the phase of the scattered field.
\end{abstract}

\maketitle

\newpage

\section{Introduction}
The field of Floquet metamaterials, also known as time-modulated materials or spacetime materials, is of considerable recent interest.  The underlying physical mechanism is to break time-reversal symmetry, by using a medium with a time-dependent refractive index or scattering potential. Notable phenomena, such as frequency conversion and parametric amplification, arise as a consequence of broken energy conservation \cite{holberg1966parametric,koutserimpas2018parametric}. A compelling class of systems feature ``travelling-wave''-like modulation, which is a natural setting for non-reciprocal wave propagation, and have recently been shown to satisfy generalized energy conservation relations \cite{zhang2024conservation,liberal2024spatiotemporal,zhang2023generalized}. Similarly, generalized energy estimates for wave equations with variable wave speeds typically feature terms which grow exponentially, enabling parametric amplification \cite{hirosawa2009generalised, ebert2015energy, cullen1958travelling, raiford1974degenerate,koufidis2024enhanced,koutserimpas2022parametric}. For scattering from a compactly supported potential, this may be understood as an instability of the system. Scattering resonances, which typically lie in the complex lower half-plane, may cross the real axis and attain a positive imaginary part \cite{demkowicz2024stability,yerezhep2021approximate,koutserimpas2018parametric, hiltunen2024coupled,ammari2023transmission}. The associated quasinormal modes are  exponentially growing in time, and the energy input exceeds the radiative losses into the far-field. For nonlinear systems, energy-preserving mechanisms for time-modulated materials have been proposed to overcome such instabilities \cite{deshmukh2022energy}.

An ideal time-dependent material only violates energy conservation when a wave field is present \cite{horsley2023quantum}. When illuminated by an incident  field, the scattered field is no longer governed by the standard optical theorem, and the material may inject or absorb energy into the scattered field. In this work, we derive a generalized energy conservation law and optical theorem for such scattering systems. We show that the classical energy conservation law must be supplemented with a source term which accounts for the energy balance of the system. Moreover, we demonstrate that the same system can have energy gain, energy loss, or energy conservation, depending on the wave harmonics present in the incident field. This may be understood as an energy imbalance due to frequency conversion, and a conserved quantity arises by redefining  the energy current associated to each frequency harmonic. Notably, we consider stable systems. The energy gain we observe is not due to parametric amplification, and the resulting field does not grow exponentially at large times.

We demonstrate the applicability of our optical theorem for imaging of time-modulated systems, without the need to measure the phase of the field. Reconstruction without phase retrieval, based on generalizations of the optical theorem, has been studied for near-field imaging of static systems \cite{carney2001near,govyadinov2009phaseless,carney1999optical}. In the weak-scattering regime, it is only possible to reconstruct the absorption of the material coefficients of static systems. In contrast, the broken energy conservation of time-modulated systems enables reconstruction of the time-dependent material coefficients.  However, the discrepancy between the probing frequency and the measured frequency (caused by frequency conversion) introduces a lower band-limit in addition to the usual upper band limit. The upper band limit behaves analogously to the  diffraction limit and restricts the image reconstruction to short length scales. In contrast, the lower band limit causes spatial characteristics of long length scales to be unresolvable. 

This paper is organized as follows. In Section \ref{sec:static}, we consider energy conservation of the  homogeneous wave equation. The setting is standard, and these results are foundational for the remainder of the work. In Section \ref{sec:eq}, we show how the time-modulated model we consider can be derived from first principles.  In Section \ref{sec:energy}, we derive a generalized energy conservation law for this model, containing a source term which accounts for energy gain or absorption. In Section \ref{sec:optical}, we formulate the optical theorem for time-modulated systems and present a detailed analysis of the energy balance of point scatterers. In Section \ref{sec:born}, we consider the weak-scattering approximation, and show how the optical theorem can be used as basis for image reconstruction based on measurements of the wave power. In Section \ref{sec:near}, we improve the reconstruction by considering near-field microscopy. Notably, the lower band-limit present in the linear inverse scattering problem may be overcome by introducing a probe in the near-field of the sample. We end the paper with some concluding remarks in Section \ref{sec:concl}.

\section{Energy density conservation for the scalar wave equation}\label{sec:static}
We begin by summarizing a few results, which will be used later on, concerning the scalar wave equation in the absence of time-dependent scatterers. This setting is standard, and may be found in  \cite{carminati2021principles,born2013principles}. We consider the wave equation with wave speed $c_0$,
\begin{equation}
	\frac{\p^2}{\p t^2}u(\bx,t) - c_0^2\Delta u(\bx,t) = 0.
\end{equation}
We let the energy density $E(\bx,t)$ be defined as 
\begin{equation}
	E(\bx,t) = \frac{1}{2}\left( \frac{1}{c_0^2}\left|\frac{\partial u}{\partial t}\right|^2 + \left|\nabla u\right|^2 \right).
\end{equation}
This energy density is associated to the energy density current $\mathbf{S}(\bx,t)$ given by
\begin{equation}
	\mathbf{S}(\bx,t) = -\Re\left(\frac{\partial u^*}{\partial t} \nabla u\right).
\end{equation}
Here, and throughout, we use $^*$ to denote the complex conjugate or conjugate transpose. $E$ and $\bS$ obey the classical energy conservation law
\begin{equation}\label{eq:Sstatic}
	\frac{\partial E}{\partial t} = -\nabla \cdot \bS.
\end{equation} If we consider a wave composed of multiple frequency harmonics centered at some quasifrequency $\omega \in \mathbb{R}$, \ie, 
\begin{equation}
	u(\bx,t) = \sum_{m=-\infty}^\infty v_m(\bx) e^{-\iu (\omega+m \Omega) t},
\end{equation}
we may pose the conservation law in the frequency domain. By time-averaging $\mathbf{S}$ over one period $T={2\pi}/{\Omega}$, we obtain
\begin{equation}
	\langle \mathbf{S}(\bx,t) \rangle = c_0\sum_{m=-\infty}^\infty \mathbf{J}_m(\bx) ,
\end{equation} 
where
\begin{equation}\label{eq:current}
\mathbf{J}_m = \frac{k_m}{2\iu} \left(v_m^*\nabla v_m - v_m \nabla v_m^* \right), \qquad 
	\bJ = \sum_{m=-\infty}^\infty \bJ_m,
\end{equation}
and we have defined $k_m = (\omega+m\Omega)/{c_0}$. In the frequency domain, we then have the conservation law
\begin{equation}\label{eq:Jstatic}
	\nabla \cdot \bJ = 0.
\end{equation}

\section{Time-modulated acoustic wave equation}\label{sec:eq}
We now define the time-modulated model which will  be studied in the remainder of this work. We consider the acoustic wave equation with time-varying wave speed $c$ which has appeared, for example, in \cite{fleury2016floquet}. To define the model, we start with the linearized Euler equation and mass conservation,
\begin{equation}
	\begin{cases}
		\ds \rho_0 \frac{\partial \mathbf{v}}{\partial t} = -\nabla p_1, \\
		\ds \frac{\partial \rho_1}{\partial t} + \rho_0 \nabla \cdot \mathbf{v} = 0.	
	\end{cases}
\end{equation}
Here,  $\mathbf{v}$ is the velocity field of the acoustic medium, $\rho_0$ is the equilibrium density while $p_1$ and $\rho_1$, respectively, are the pressure and density fluctuations due to the acoustic wave.  Moreover, we consider the linearized equation of state
\begin{equation}
	p_1(\bx,t) = c^2 \rho_1(\bx,t).
\end{equation}
Here, $c$ is the speed of acoustic waves. We consider a situation whereby $c = c(\bx,t)$ is periodically modulated in time. To preserve mass balance, we take a constant $\rho_0$. In this case, we obtain the time-modulated wave equation
\begin{equation}\label{eq:wavemodel}
	\frac{\partial^2}{\partial t^2}\left(\frac{1}{c^2(\bx,t)}p_1(\bx,t) \right) - \Delta p_1(\bx,t) = 0.
\end{equation}
Equation \eqref{eq:wavemodel} describes scalar waves with a time-modulated refractive index: For a reference wave speed $c_0$ (constant in $t$ and $\bx$) we let the refractive index $n(\bx,t)$ be defined through
\begin{equation}
	c(\bx,t) = \frac{c_0}{n(\bx,t)}.
\end{equation}

\section{Generalized energy conservation}\label{sec:energy}
We now turn to the problem with time-dependent scatterers and derive a generalized energy conservation. We consider the equation
\begin{equation}\label{eq:wave}
	\frac{\p^2}{\p t^2} \left(\frac{n^2(\bx,t)}{c_0^2}u(\bx,t)\right) - \Delta u(\bx,t) = 0, \end{equation}
where $n(\bx,t)$ is $T$-periodic in $t$. By Bloch's theorem in $t$, we seek solutions which are quasiperiodic in $t$:
\begin{equation}
	u(\bx,t) = \sum_{m=-\infty}^\infty v_m(\bx) e^{-\iu (\omega+m\Omega)t},
\end{equation}
where $\Omega = {2\pi}/{T}$.  We let $\eta$ be the perturbation of $n^2$ against a (constant) background:
\begin{equation}
	n^2(\bx,t) = 1+\eta(\bx,t).
\end{equation}
We consider a general (possibly complex-valued) spacetime-varying $\eta(\bx,t)$  with  Fourier coefficients $\eta_m(\bx)$, \ie,
\begin{equation}\label{eq:fourier_eta}\eta(\bx,t) = \sum_{m=-\infty}^\infty \eta_m(\bx) e^{-\iu m\Omega t}.\end{equation}
Substituting into \eqref{eq:wave} results in a system of coupled equations
\begin{equation}\label{eq:spatial}
	\begin{split}
		&\Delta v_m + k_m^2v_m +  k_m^2\sum_{l=-\infty}^\infty \eta_{m-l}(\bx) v_l  =0,
	\end{split}
\end{equation}
where, as before, $k_m=(\omega+m\Omega)/{c_0}$. We begin by writing \eqref{eq:spatial} in matrix form: define the (doubly infinite) Toeplitz matrix  $H = H(\bx)$ and the diagonal matrix $K$ as 
\begin{equation}
	H = \begin{psmallmatrix} \ddots& \ddots & \ddots  &  & \hspace{-5pt}   \\
		\ddots& \eta_0 & \eta_{-1} & \eta_{-2}&  \\
		\ddots & \eta_{1} & \eta_0 & \eta_{-1}  & \ddots \\
		& \eta_{2} & \eta_{1} & \eta_0  & \ddots \\
		&  & \ddots  & \ddots & \ddots \end{psmallmatrix}, \quad
	K = \begin{psmallmatrix} \ddots \hspace{-3pt}&  &   &  &    \\
		& k_{-1} \hspace{-5pt} & &  &   \\
		& & k_{0}\hspace{-2pt} & &  \\
		& & & k_1 &  \\[-0.3em]
		&  &   &  & \hspace{-3pt}\ddots \end{psmallmatrix}.
\end{equation}	
Moreover, let $\sv(\bx) = \begin{psmallmatrix} \vdots \\[-0.1em] v_{-1}\\v_0\\v_1\\[-0.2em] \vdots \end{psmallmatrix}$.
We can then write \eqref{eq:spatial} as 
\begin{equation}\label{eq:matrixform}
	\Delta \sv(\bx) + K^2\sv(\bx) + K^2H(\bx)\sv(\bx) = 0.
\end{equation}

We next phrase the conservation law of the energy density current \eqref{eq:current} when applied to the spacetime-varying material defined by  $\eta(\bx,t)$. We let $\langle \cdot, \cdot \rangle$ denote the $\ell^2$-inner product in the frequency variable. Taking the inner product of \eqref{eq:matrixform} by $K\sv$ gives 
\begin{equation}\label{eq:mult}
	\langle K\sv, \Delta \sv\rangle + \langle K\sv,  K^2\sv\rangle + \langle K\sv, K^2H\sv\rangle = 0.
\end{equation}
We will subtract the above equation by its conjugate. Using the identity $\overline{f}\Delta f - f \Delta \overline{f} = \nabla\cdot\left(\overline{f}\nabla f - f \nabla \overline{f} \right)$, we find that 
\begin{equation}\label{eq:inter}
	\frac{1}{2\iu}\bigl(\langle  K \sv, \Delta \sv\rangle - \langle  K {\sv}, \Delta {\sv}\rangle^*\bigr) = \nabla \cdot \sum_{i=-\infty}^\infty \mathbf{J}_i,
\end{equation}
where the energy current $\mathbf{J}_i$ of the $i$\textsuperscript{th} harmonic $v_i$ is defined as in \eqref{eq:current}. We also observe that 
\begin{equation}
\frac{1}{2\iu}\left(\langle K \sv, K^2H\sv\rangle-  \langle K {\sv}, K^2{H}{\sv}\rangle^*\right)= \langle \sv, M\sv \rangle,
\end{equation}
where the matrix $M$ is defined as $M = \frac{1}{2\iu}\left( K^3H - \left( K^3H\right)^*\right)$. Together with \eqref{eq:mult} and \eqref{eq:inter}, we find the conservation law
\begin{equation}\label{eq:cons}
\nabla \cdot \bJ +  \langle \sv, M\sv \rangle = 0.
\end{equation}
Eq. \eqref{eq:cons} is the generalized energy conservation of the time-modulated system, and generalizes \eqref{eq:Jstatic}. Note that we do not need to assume that $\eta(\bx,t)$ is compactly supported in $\bx$ for \eqref{eq:cons} to hold, although this will be the main case considered subsequently. 

Observe that, even in the case when $\eta(\bx,t)$ is real-valued (whereby $H$ is self-adjoint), the matrix $ K^3H$ is in general not self-adjoint, and $M$ defines a source term for the conservation law. If $\eta(\bx,t)$ is real-valued and constant in $t$, \eqref{eq:cons} reduces to the energy conservation $\nabla \cdot \bJ = 0$ for static materials. We can phrase this conservation law in the time domain:
\begin{equation}
	\frac{\partial E}{\partial t} = -\nabla \cdot \bS -\frac{1}{2c_0^2} \Re\left( \eta\frac{\partial \overline{u}}{\partial t}\frac{\partial^2 u}{\partial t^2}\right),
\end{equation}
which generalizes \eqref{eq:Sstatic}. Here, the source term $\Re\left( \eta\frac{\partial \overline{u}}{\partial t}\frac{\partial^2 u}{\partial t^2}\right)$ corresponds to the operator $M$ in frequency domain.

We note that \eqref{eq:cons} is phrased for the conventional energy current $\bJ_i$ defined as in \eqref{eq:current}. For nonabsorbing media, the source term can be eliminated by redefining the weights of the energy current of each frequency harmonic. We define the generalized energy current $\bm{\mathcal{J}}_i$ as 
\begin{equation}\label{eq:currenthat}
	\bm{\mathcal{J}}_i = \frac{1}{2\iu k_i^2} \left(\overline{v_i}\nabla v_i - v_i \nabla \overline{v_i} \right).
\end{equation}
The conservation law is now given by 
\begin{equation}\label{eq:conshat}
	\nabla \cdot \bm{\mathcal{J}} + \langle \sv, \mathcal{M}\sv \rangle = 0,
\end{equation}
for $\mathcal{M} = \frac{1}{2\iu}\left(H-H^*\right).$ In particular, for real-valued $\eta(\bx,t)$ we have $\mathcal{M} = 0$ and $\nabla \cdot \bm{\mathcal{J}} = 0$.

\section{Optical theorem}\label{sec:optical}
\begin{figure}[tbh]
	\centering
	\includegraphics{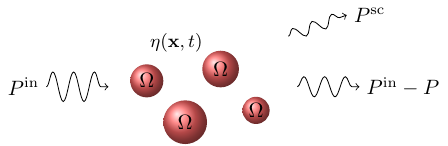}
	\caption{Schematic illustration of the scattering experiment. An incident field to a time-modulated system gives rise to a scattered field of power $\Psc$. In the case of energy conservation, $\Psc$ must agree with the extinguished power $P$ which is depleted from the incident field. Energy gain corresponds to the case $\Psc > P$ while energy absorption corresponds to $\Psc < P$.}\label{fig:sketch}
\end{figure}

We now turn to scattering from compactly supported $\eta(\bx,t)$. When formulating the optical theorem for scattering from a compactly supported domain we will consider the conventional current $\bJ$ defined as in \eqref{eq:current}; the active energy input/output will be determined through an energy balance in the exterior of the scattering domain. We illustrate the setup and notation in Fig. \ref{fig:sketch}.

\subsection{Computation of the active and total power}
As before, we define the total energy current $\bJ$ as 
\begin{equation}\label{eq:J}
\bJ = \sum_{i=-\infty}^\infty \frac{k_i}{2\iu} \left(\overline{v_i}\nabla v_i - v_i \nabla \overline{v_i} \right).
\end{equation}
We separate the field $v_i$ into incident and scattered parts:
\begin{equation}\label{eq:v}
	v_i = \vin_i+\vsc_i,
\end{equation}
and define the incident and scattered energy currents $\Jin$ and $\Jsc$ analogously to \eqref{eq:J}, but for each wave component separately. We define the incident, scattered, and active power, respectively, as
\begin{equation}\label{eq:pdef}
\Pin = \int_{\partial D} \Jin \cdot \hat{\bn} \dx \bx, \quad 
\Psc = \int_{\partial D} \Jsc \cdot \hat{\bn} \dx \bx, \quad 
\Pop = -\int_{\partial D} \bJ \cdot \hat{\bn} \dx \bx,
\end{equation}
where $D$ is the volume of the scatterer. The active power describes the energy input or output from the time-varying material. In the static case, this describes the absorbed energy and we necessarily have $\Pop \geq 0$. For time-varying materials, the sign of $\Pop$ is no longer fixed, and we may have modes of operation with energy input, energy output, or energy conservation. From \eqref{eq:cons}, we have 
\begin{equation}\label{eq:Pact}
	\Pop = \Im\int_{D} \langle \sv, K^3H\sv \rangle \dx \bx.
\end{equation} 
We define the total power $\Pext$ by
\begin{equation}
\Pext = \Psc + \Pop.
\end{equation}
In the static case, $P$ is the power which is extinguished from the total field in the forward direction of the incident field due to the scatterer, and we have $\Pext\geq 0$. For time-dependent scatterers, the sign of $\Pext$ is no longer fixed, and the scatterers may instead inject energy into the incident field.

For plane-wave incidence, it is straightforward to show that $\Pin = 0$. From the definitions in \eqref{eq:v} and \eqref{eq:pdef}, along with \eqref{eq:Pact}, it then follows that the scattered and total power are given by
\begin{equation}\label{eq:Ps}
	\Psc = -\Im\int_{D} \langle \vsc, K^3H\sv \rangle \dx \bx, \quad P = \Im\int_{D} \langle \vin, K^3H\sv \rangle \dx \bx.
\end{equation}

We consider a general incident field consisting of multiple frequency components, each composed of a superposition of plane waves:
\begin{equation}\label{eq:vin}
	\vin_i(\bx) = \int a_i(\hat{\bz})e^{\iu k_i \hat{\bz}\cdot \bx}\dx \hat{\bz},
\end{equation}
for some complex amplitudes $a_i(\hat{\bz})$, $i\in \Z$. We will use the superposition principle to express the scattered field $\vsc$. To this end, we fix $j\in \Z$ and let $\esc_{ij}(\bx,\hat{\bz})$ be the scattered field given the incident field $\ein_{ij}(\bx,\hat{\bz})$ defined by
\begin{equation}
	\ein_{ij}(\bx,\hat{\bz}) = \delta_{ij}a_i(\hat{\bz})e^{\iu k_i \hat{\bz}\cdot \bx}.
\end{equation}
From the system \eqref{eq:spatial} of coupled equations, the Lippmann-Schwinger equation is of the form
\begin{equation}\label{eq:LS}
	\vsc_i(\bx) = k_i^2\sum_{j=-\infty}^\infty\int_D G^{k_i}(\bx-\by)\eta_{i-j}(\by) v_j(\by) \dx \by,
\end{equation}
where $G^{k}(\bx)$ is the outgoing Green's function for the Helmholtz equation,
\begin{equation}
G^{k}(\bx) = \frac{e^{\iu k |\bx|}}{4\pi|\bx|}.
\end{equation} 
This means that $\esc$ has a far-field expansion given by 
\begin{equation}
	\esc_{ij}(\bx,\hat{\bz}) \sim 4\pi G^{k_i}(\bx) A_{ij}(\hat{\bx},\hat{\bz}),
\end{equation}
where $\sim$ denotes equality up to $O(|\bx|^{-2})$ and the scattering amplitudes $A_{ij}(\hat{\bx},\hat{\bz})$ are given by
\begin{equation}
A_{ij}(\hat{\bx},\hat{\bz}) = \frac{k_{i}^2}{4\pi}\sum_{j'=-\infty}^\infty \int_D e^{-\iu k_i\hat{\bx}\cdot \by}\eta_{i-j'}(\by)e_{j'j}(\by,\hat{\bz}) \dx \by,
\end{equation}
for $e_{ij}(\bx,\hat{\bz}) = \ein_{ij}(\bx,\hat{\bz}) + \esc_{ij}(\bx,\hat{\bz})$. 
Here, $A_{ij}(\hat{\bx},\hat{\bz})$ is the scattering amplitude of the $i$\textsuperscript{th} harmonic in direction $\hat{\bx}$  for a given incident harmonic $j$ in direction $\hat{\bz}$. By linearity, it now follows that $\vsc$ satisfies the far-field expansion
\begin{equation}\label{eq:FF}
	\vsc_i(\bx) \sim 4\pi G^{k_i}(\bx) \int \sum_{j=-\infty}^\infty A_{ij}(\hat{\bx},\hat{\bz})a_j(\hat{\bz})\dx \hat{\bz}.
\end{equation}
Comparing \eqref{eq:LS} with \eqref{eq:FF} we find the following relation, which will be useful subsequently,
\begin{equation}\label{eq:Aa}
	\int \sum_{j=-\infty}^\infty A_{ij}(\hat{\bx},\hat{\bz})a_j(\hat{\bz})\dx \hat{\bz} = \frac{k_i^2}{4\pi}\sum_{j=-\infty}^\infty\int_D e^{-\iu k_i\hat\bx\cdot \by} \eta_{i-j}(\by) v_j(\by) \dx \by.
\end{equation}

It is natural to introduce the space $\L$ of $\ell^2$-sequences of $L^2$-functions on $S^2$. Note that the scattering amplitudes  define a matrix-integral operator $\cA$ on $\L$ by
\begin{equation}
	(\cA f)_i(\hat{\bx}) = \int \sum_{j=-\infty}^\infty A_{ij}(\hat{\bx},\hat{\bz})f_j(\hat{\bz})\dx \hat{\bz} .
\end{equation}
We also define the inner product $\langle\cdot , \cdot \rangle_\L$ on $\L$ as
\begin{equation}
	\langle f, g\rangle_\L = \int \sum_{i=-\infty}^\infty \overline{f_i}(\hat{\bx}) g_i(\hat{\bx}) \dx \hat{\bx},
\end{equation}
and let $\| \cdot \|_\L$ denote the associated norm. A direct calculation now yields
\begin{equation}
	\Psc = \int  \sum_{i=-\infty}^\infty k_i^2 \left| \int \sum_{j=-\infty}^\infty A_{ij}(\hat{\bx},\hat{\bz})a_j(\hat{\bz}) \dx \hat{\bz}\right|^2\dx \hat{\bx},\label{eq:psc}
\end{equation} 
which can concisely be written $\Psc = \| K \cA \ca\|_\L^2$.

We next compute the total power $\Pext$. Combining \eqref{eq:Ps} with \eqref{eq:Aa} gives
\begin{equation}
\Pext = 4\pi \Im\int \int \sum_{i,j=-\infty}^\infty k_i \overline{a_i}(\hat{\bx})A_{ij}(\hat{\bx},\hat{\bz}) a_j(\hat{\bz}) \dx \hat{\bx}\dx \hat{\bz}.
\end{equation}
In terms of the inner product in $\L$, this can be written as
\begin{equation}\label{eq:optical}
	\Pext =4\pi \Im \langle \ca,  K\cA\ca\rangle_{\L}.
\end{equation}
Equation \eqref{eq:optical} is the optical theorem for time-modulated scatterers, describing the total power in terms of the scattering coefficients $A_{ij}$ of the scatterer. Note that it is a quadratic in the vector $\ca$, which describes the relative amplitudes and phases of the incident harmonics.

Combining \eqref{eq:psc} and \eqref{eq:optical}, we can write the active power $\Pop$ as
\begin{equation}\label{eq:pop}
	\Pop = 4\pi \Im \langle \ca,  K\cA \ca\rangle_\L - \left\|   K \cA\ca \right\|^2_\L.
\end{equation}
Note that $\Pop = 0$ corresponds to an energy-neutral scatterer without absorption or gain. 

We can repeat the analysis but based on the generalized energy current defined in \eqref{eq:currenthat}. We assume that $\Im \bigl(\eta(x,t)\bigr) = 0$, whereby the source term of \eqref{eq:conshat} vanishes. Following the same steps as before, we have
\begin{equation}
	4\pi \Im \langle \ca,  K^{-2}\cA \ca\rangle_\L - \langle \cA\ca,   K^{-1} \cA\ca \rangle_\L = 0,
\end{equation}
for any $\ca$. In particular, if $\ca$ is an eigenstate of $A$ corresponding to eigenvalue $\lambda$, we have 
\begin{equation} \label{eq:lambda}
	4\pi \Im \lambda = |\lambda|^2\frac{\langle \ca,   K^{-1} \ca \rangle_\L}{\| K^{-1} \ca\|^{2}}.
\end{equation}

\subsection{Point-particle scattering}
\begin{figure*}[tbh]
	\centering
	\includegraphics[width=\linewidth]{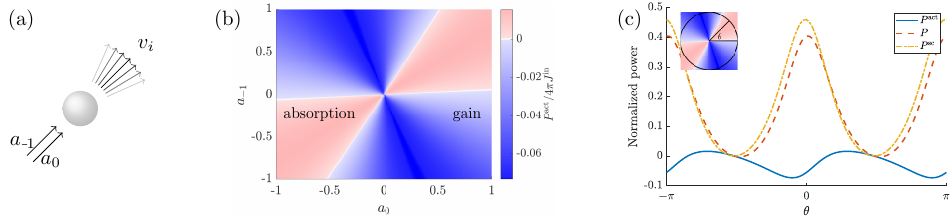}
	\caption{Energy phase diagram of $\Pop$ for a single time-dependent point scatterer. We consider a situation with two incident harmonics $n=0$ and $n=-1$ as sketched in (a). Along the coordinate axes of (b), where one of the incident harmonics vanishes, we have $\Pop <0$ and the total system has energy gain. When both harmonics are present, there are regimes of energy absorption, energy conservation, and energy gain. In (c), we plot the active, total, and scattered power, respectively, where the active power is given by the difference of the other two.
	} \label{fig:energy_phase}
\end{figure*}

We will illustrate the previous analysis in the case of small, high-contrast particles which has previously been studied in  \cite{hiltunen2024coupled}. In the case of a single scatterer, we let $D=\epsilon B + \by$ for some fixed domain $B$ and $0<\epsilon \ll 1$ and assume that $\eta$ is proportional to $\epsilon^{-2}$:
\begin{equation}
	\eta(\bx,t) = \frac{s(t)}{\epsilon^2}, \quad \bx\in D,
\end{equation}
for some function $s(t)$ independent of $\epsilon$. In this case, it has been shown in \cite{hiltunen2024coupled} that the scattered field is of order $O(1)$ as $\epsilon \to 0$. Moreover, the particle behaves as a point scatterer in the sense that the scattered field is given by a point source at the location $\by$ of the particle,
\begin{equation}
	\vsc_i(\bx) = G^{k_i}(\bx-\by) \sum_{j=-\infty}^\infty \Lambda_{ij}\vin_j(\by),
\end{equation}
for $\bx \neq \by$ and as $\epsilon \to 0$. Here $\Lambda_{ij}$ are the (angle-independent) scattering coefficients of the particle, and are related to the far-field amplitude through
\begin{equation}
	\Lambda_{ij} = 4\pi A_{ij}(\hat{\bx},\hat{\bz}) e^{\iu k_i\hat{\bx}\cdot \by}e^{-\iu k_j \hat{\bz}\cdot \by}.
\end{equation}
We take two incident harmonics: $n=0$ and $n=-1$ and assume that the scatterer itself is non-absorbing: $\Im \bigl(\eta(\bx,t)\bigr) = 0$. In Fig.~\ref{fig:energy_phase}, we plot a phase diagram of  $\Pop$ for varying $a_0$ and $a_1$. We note that the same system may exhibit either energy gain or energy absorption depending on the incident field. These two cases are separated by phase boundaries where the energy is conserved.

\begin{figure*}[tbh]
	\centering
	\includegraphics[width=0.8\linewidth]{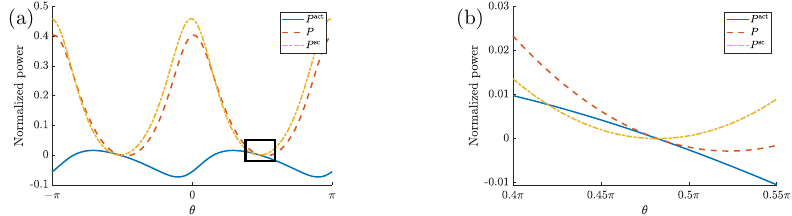}
	\caption{Since $\cA$ is low-rank, there may be points (shown in (b)) at which the scattered field vanishes. At these points, $\Pop$ changes sign, signifying a transition from energy absorption to energy gain. Close to these points, the total energy $\Pext$ may turn negative and the scatterer injects, rather than depletes, energy from the incident field.} \label{fig:psc_pop}
\end{figure*}

We note that the matrix $\cA$ can have low rank. For example, in the point-scattering approximation of \cite{hiltunen2024coupled}, $\cA$ has rank $1$ close to a simple resonance. Zeros of $\Pop$ therefore appear either when the two terms of \eqref{eq:pop} cancel, or when $\ca$ is a kernel vector of $\cA$. The former case can be viewed as a generalisation of the energy conservation of a non-absorbing time-invariant scatterer, where the extinguished power matches the scattered power. In the latter case, the incident field does not couple to the scatterer: The scattered field vanishes and $\Pop=\Pext=\Psc=0$ as highlighted in Fig.~\ref{fig:psc_pop}. Around this region, the total power may turn negative and the scatterer amplifies the total field in the forward direction of the incident wave.

It is worth emphasizing that the exhibited energy gain does not originate from instability of the system: The imaginary part of the resonant quasifrequencies remain negative and the solution does not exhibit an exponential growth as $t\to \infty$.

\begin{figure*}[tbh]
	\centering
	\includegraphics[width=1\linewidth]{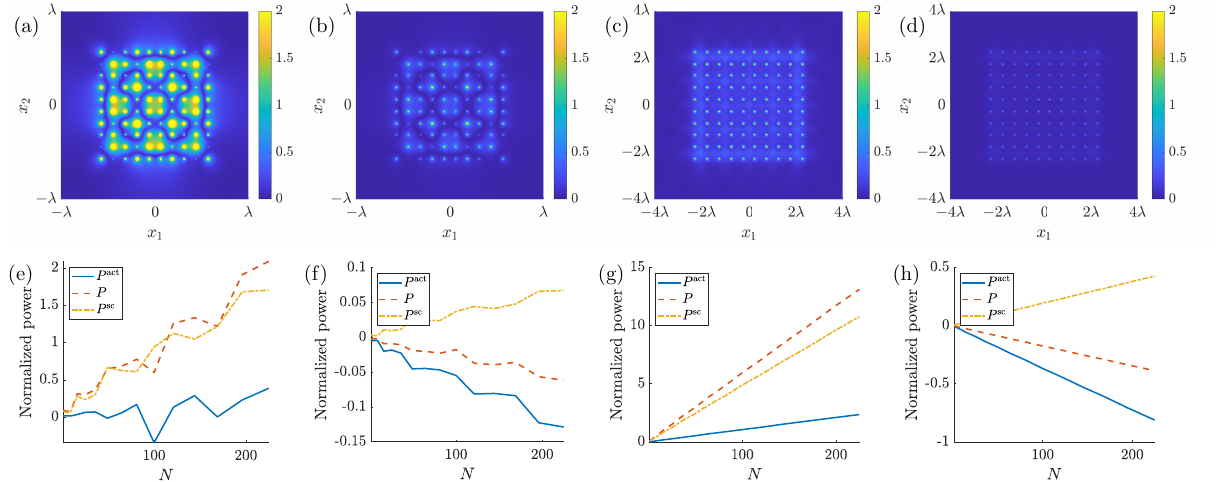}
	\caption{We plot examples of the scattered field  ((a)--(d)) in a square configuration of scatterers, and corresponding normalized power ((e)--(h)) as a function of the number $N$ of particles, for two different values of $\theta$ as defined in Fig.~\ref{fig:energy_phase}. In (a),(b),(e),(f), we consider closely spaced scatterers (relative to the wavelength of the dominant harmonic). In this case, scattered modes hybridize and the collective response may be different compared to an isolated particle. For well-separated particles, the hybridization is negligible and the normalized powers $\Pop$, $\Pext$, and $\Psc$ are directly proportional to the number of particles.} \label{fig:sc_N}
\end{figure*}

For multiple point scatterers at positions $\by_m$, we use a fully-coupled approach to compute the scattered field. We take $N$ small particles defined by 
\begin{equation}D = \bigcup_{m=1}^N \epsilon B + \by_m.\end{equation}
The scattered field is then given by 
\begin{equation}
	\vsc_i(\bx)=  \sum_{m=1}^N  G^{k_i}(\bx-\by_m) \sum_{j=-\infty}^\infty \Lambda_{ij}\alpha_{jm},
\end{equation}
where $\alpha_{jm}$ are the solutions of the fully-coupled linear system 
\begin{equation}
	\alpha_{im} = \vin_i(\by_m) + \sum_{n\neq m}G^{k_i}(\by_m-\by_n)\sum_{j=-\infty}^\infty \Lambda_{ij}\alpha_{jn}.
\end{equation}	
Computing the scattered and total power based on \eqref{eq:psc} and \eqref{eq:optical}, we then have
\begin{equation}
	\Psc = \sum_{i=-\infty}^\infty k_i \sum_{m,n=1}^N\overline{c_{im}}c_{in}\frac{\sin(k_i|\by_n-\by_m|)}{4\pi|\by_n-\by_m|},
\end{equation}
and 
\begin{equation}
	\Pext = \Im \sum_{i=-\infty}^\infty k_i \sum_{m=1}^N (\vin_i)^*(\by_m)c_{im},
\end{equation}
where we use the shorthand $c_{im} =  \sum\limits_{j=-\infty}^\infty \Lambda_{ij}\alpha_{jm}$.
We illustrate these results in Fig.~\ref{fig:sc_N}, where we plot the scattered field in a square configuration of point particles. We consider two values of $\theta$ as defined in Fig.~\ref{fig:energy_phase}(c), corresponding, respectively, to the cases $\Pop, \Pext >0$ and $\Pop, \Pext < 0$. In the cases $\Pext <0$, we note that $\Psc$ is close to zero, and the overall scattered has a low amplitude. In each case,  the energy gain or absorption is amplified by including more resonators and is, in the case of well-separated resonators, directly proportional to the number of particles. 

\section{Linear inverse scattering problem}\label{sec:born}
We next consider linearized inverse scattering and the reconstruction of the time-dependent material coefficient $\eta$. The reconstruction is based on measurements of the power $\Pext$ and does not require measurements of the phases of the waves.

\subsection{Weak scattering approximation}
We consider the perturbative regime of small $\eta$, in which we can adopt the Born approximation of the scattered field. The scattered field is then given by
\begin{equation}
	\vsc_i(\bx) = k_i^2\int \sum_{j=-\infty}^\infty  a_j(\hat{\bz})\int_D G^{k_i}(\bx-\by)\eta_{i-j}(\by)e^{\iu k_j \hat{\bz}\cdot \by} \dx \hat{\bz} \dx \by.
\end{equation}
Similarly, the Born approximation of the scattering amplitude $A$ is 
\begin{equation}\label{eq:Aborn}
	A_{ij}(\hat{\bx},\hat{\bz}) = \frac{k_i^2}{4\pi}\int_D e^{-\iu \left(k_i\hat{\bx} - k_j\hat{\bz}\right)\cdot \by}\eta_{i-j}(\by) \dx \by,
\end{equation}
so that $A_{ij}(\hat{\bx},\hat{\bz})$ is given by the Fourier transform of $\eta_{i-j}$:
\begin{equation}
	A_{ij}(\hat{\bx},\hat{\bz}) = \frac{k_i^2}{4\pi} \widetilde{\eta}_{i-j}(k_i\hat{\bx} - k_j\hat{\bz}).
\end{equation}

\subsection{Reconstruction without phase retrieval}
Within the weak scattering approximation, we cannot reconstruct the real part of $\eta_0$ without phase retrieval (by solely measuring $\Pext$). Nevertheless, as will be shown in this section, we are able to reconstruct $\eta_i$, for $i\neq 0$, based on \eqref{eq:optical}. This will be made possible by taking an incident field composed of two plane waves of directions $\hat{\bz}_1, \hat{\bz}_2$ and with distinct frequencies $k_{j_1}$ and $k_{j_2}$:
\begin{equation}
	\vin_i(\bx) = c_1\delta_{ij_1} e^{\iu \bk_1\cdot \bx} + c_2\delta_{ij_2} e^{\iu \bk_2 \cdot \bx},
\end{equation}
where $\bk_n = k_{j_n} \hat{\bz}_n$ for $n=1,2$. This gives
\begin{equation}\label{eq:pext_c}
	\Pext(c_1,c_2) = \bc^* M \bc,
\end{equation}
where $\bc = \begin{psmallmatrix}c_1\\c_2\end{psmallmatrix}$, $Q = \frac{1}{2\iu}\left(Q-Q^*\right)$, and the matrix elements of $Q$ are given by
\begin{align}
	Q_{11} &= k_{j_1}^3\int_D \eta_0(\by) \dx \by, & Q_{12} &=  k_{j_1}^3\widetilde{\eta}_{j_1-j_2}(\bk_1 - \bk_2), \\ Q_{21} &=  k_{j_2}^3 \widetilde{\eta}_{j_2-j_1}(\bk_2 - \bk_1),  &  Q_{22} &= k_{j_2}^3\int_D \eta_0(\by) \dx \by.
\end{align}
We consider the case $\Im \bigl( \eta(\bx,t)\bigr) = 0$ and seek to reconstruct $\eta_i(\bx)$ for $i\neq 0$. We observe that $\widetilde{\eta}_i(\bk) = \overline{\widetilde{\eta}_{-i}(-\bk)}$. In this case, the off-diagonal term $\widetilde{\eta}_{j_1-j_2}(\bk_1 - \bk_2)$ can be reconstructed in the case $j_1\neq j_2$ using two measurements of $\Pext$:
\begin{equation}\label{eq:eta}
	\widetilde{\eta}_{j_1-j_2}(\bk_1 - \bk_2) = \frac{1}{k_{j_1}^3-k_{j_2}^2}\bigl(\Pext(1,\iu) + \iu \Pext(1,1)\bigr).
\end{equation} 
We fix $j_1$ and $j_2$, and define the data function $D(\hat{\bz}_1,\hat{\bz}_2)$ as follows:
\begin{equation}\label{eq:D1}
	 D(\hat{\bz}_1,\hat{\bz}_2) = \frac{1}{k_{j_1}^3-k_{j_2}^3}\bigl(\Pext(1,\iu) + \iu \Pext(1,1)\bigr).
\end{equation}
Equation \eqref{eq:eta} then becomes
\begin{equation}\label{eq:D2}
	D(\hat{\bz}_1,\hat{\bz}_2) = \int \eta_{j_1-j_2}(\bx)e^{-\iu(\bk_1 - \bk_2)\cdot \bx} \dx \bx.
\end{equation}
Equation \eqref{eq:D2} gives band-limited data of $\eta_{j_1-j_2}$ and provides a means of (resolution-limited) reconstruction. We note that it is not necessary to measure the phases of the waves. In light of \eqref{eq:Aborn}, knowledge of the scattering amplitude $A_{ij}(\hat{\bx},\hat{\bz})$ does not improve the reconstruction of $\eta_i(\bx)$ for $i\neq 0$. 

We consider a spherical shell of inner radius $k_r$ and outer radius $k_R$:
\begin{align}
	\left\{ k_{j_1} \hat{\bz}_1 - k_{j_2} \hat{\bz}_2 \mid \hat{\bz}_1, \hat{\bz}_2\in S^2 \right\} &= \left\{\bk  \mid k_r\leq|\bk|\leq k_R \right\}  ,
\end{align}
where 
\begin{equation}
	k_r = \Bigl| \left|k_{j_1}\right| - \left|k_{j_2}\right|\Bigr| \quad \text{and} \quad k_R = \left|k_{j_1}\right| + \left|k_{j_2}\right|.
\end{equation}
It follows that we are able to reconstruct an upper and lower band-limited approximation of $\eta_{j_1-j_2}(\bx)$:
\begin{equation}
	\eta_{j_1-j_2}(\bx) = \frac{1}{(2\pi)^3}\int_{ k_r\leq|\bk|\leq k_R} D(\bk) e^{\iu \bk \cdot \bx} \dx \bk.
\end{equation}
If both $k_{j_1}$ and $k_{j_2}$ are positive, we have  
\begin{equation}
	k_r = \frac{\left|j_1-j_2\right|\Omega}{c} \quad \text{and} \quad k_R =  \frac{2\omega + (j_1+j_2)\Omega}{c}.
\end{equation}
The upper band limit corresponds to the usual resolution limit  at the average of the two frequencies (incident and scattered) appearing in the measurement. Instead of $d= \lambda/2$, the resolution limit $d$ is now
\begin{equation}\label{eq:reslim}
	d= \frac{\lambda_1\lambda_2}{\lambda_1+\lambda_2},
\end{equation}
where $\lambda_1$ and $\lambda_2$ are the wavelengths of the two harmonics.

\begin{figure}[tbh]
	\begin{subfigure}[t]{0.33\linewidth}
		\includegraphics[width=\linewidth]{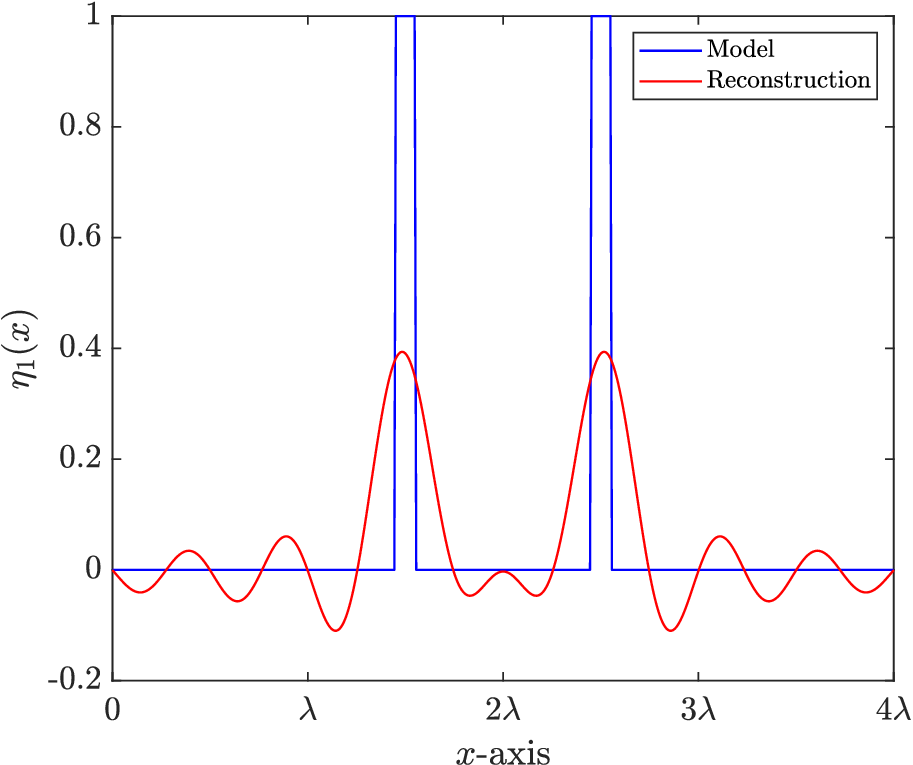}
		\caption{$\Omega\ell = 0.01$}
	\end{subfigure}\hfill
\begin{subfigure}[t]{0.33\linewidth}
\includegraphics[width=\linewidth]{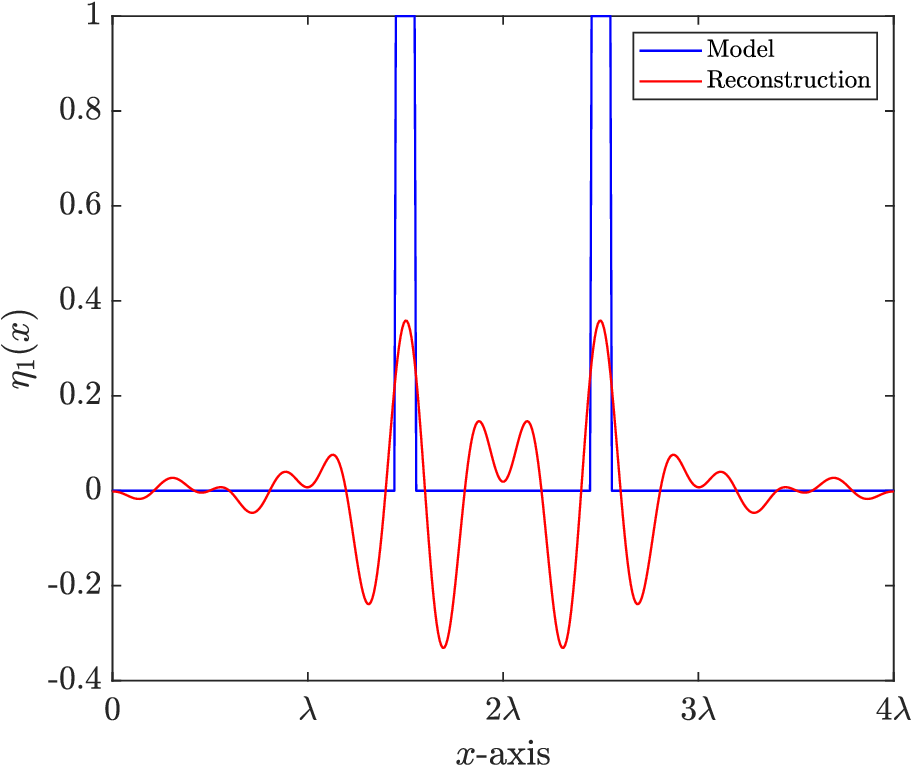}
\caption{$\Omega\ell = 1$}
\end{subfigure}\hfill
\begin{subfigure}[t]{0.33\linewidth}
\includegraphics[width=\linewidth]{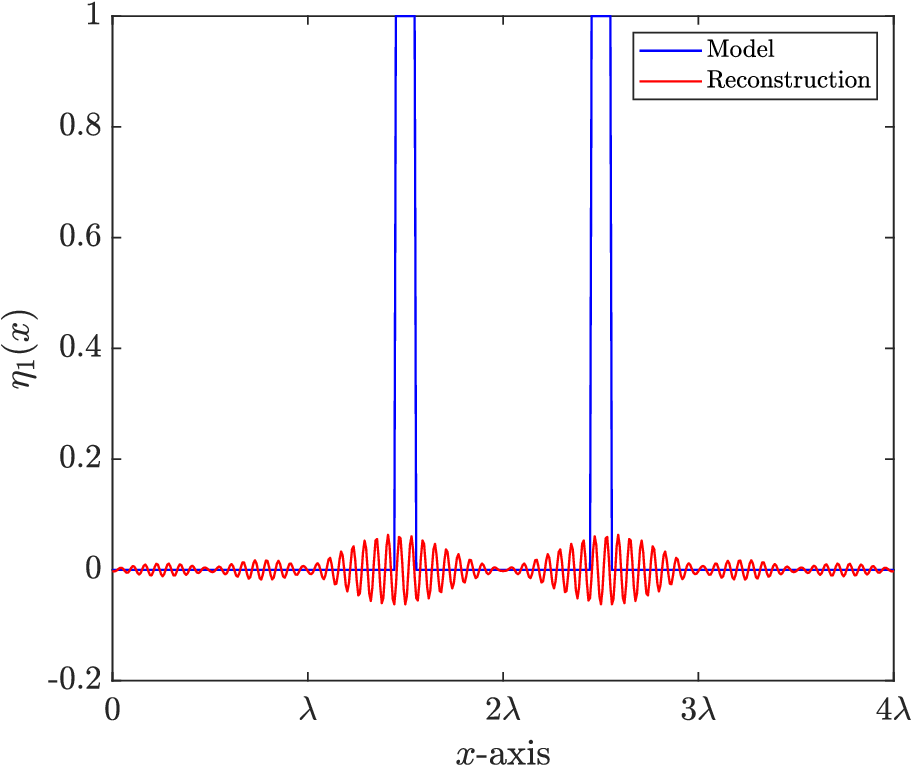}
\caption{$\Omega\ell = 10$}
\end{subfigure}
\caption{One-dimensional example of a reconstructed image of $\eta_1(\bx)$ using a model with two piecewise constant ``particles'' of width $\ell = 0.1\lambda$ and separation distance $d = \lambda$. If $\Omega\ell \ll 1$, it is possible to reconstruct the image up to the classical resolution limit given by \eqref{eq:reslim}. If $\Omega\ell \approx 1$, the reconstructed image is both upper and lower band-limited. If $\Omega\ell \gg 1$, the reconstruction fails.}\label{fig:reconstruct1D}
\end{figure}

The lower band limit depends purely on the modulation frequency $\Omega$ and not on the frequency of the incident and scattered waves. This gives an upper limit to the size of a featue that can be reconstructed. In addition, the accuracy of the reconstruction depends on the value of $\Omega$ in relation to the characteristic length scale $\ell$ of $\eta$. In the ``slow-modulation'' regime, given by $\Omega \ell \ll 1$, the lower band limit is negligible and the resolution of the reconstructed image is governed by the standard resolution limit (see Fig.~\ref{fig:reconstruct1D}(a)). On the other hand, in the ``fast-modulation'' regime, given by $\Omega\ell \gg 1$, the lower band limit will severely restrict the resolution of the reconstructed image (see Fig.~\ref{fig:reconstruct1D}(c)), and image features at large length scales are invisible in the reconstructed image. In the intermediate regime, the reconstructed image is both upper and lower band-limited, and artifacts may appear in the reconstructed image (see  Fig.~\ref{fig:reconstruct1D}(b)).

\section{Near-field inverse problem}\label{sec:near}
We seek to overcome the band limit of the reconstructed image using near-field microscopy, where a probing tip is introduced in the near-field of the sample. The interaction between the sample and the tip provides further knowledge about the sample, and provides the basis for a high-resolution reconstruction. 

We follow the approach of \cite{sun2006near,sun2007strong} and model the probe tip by a point scatterer. We consider the system
\begin{equation}\label{eq:total}
	\Delta \sv(\bx) +  K^2\sv(\bx) +  K^2H(\bx)\sv(\bx) = 0,
\end{equation}
where  $\eta_n(\bx)$ is now a sum of two terms coming from the sample $s_n(\bx)$ (which we seek to recover) and the tip $t_n(\bx)$ (which is assumed to be known):
\begin{equation}\label{eq:spt}
	\eta_n(\bx) = s_n(\bx) +t_n(\bx).
\end{equation}
We adopt a Green's function approach to solve \eqref{eq:total}. The Green's function $ G =  G(\bx,\bx',n,n')$ is given by the solution to the problem
\begin{equation}
	\Delta  G +  K^2 G +  K^2H(\bx) G = -\delta(\bx-\bx')\delta_{nn'}.
\end{equation}
Here $\delta(\bx-\bx')\delta_{nn'}$ represents a point source of the $n'${th} harmonic at position $\bx'$. Given the scattering kernel 
\begin{equation}
	V(\bx_1,\bx_2,n_1,n_2) = \delta(\bx_1-\bx_2)\eta_{n_1-n_2}(\bx_1)k_{n_1}^2, 
\end{equation}
we introduce the summation-integration operator $V$ as 
\begin{equation}
	(\cA V B)(\bx,\bx',n,n') = \sum_{n_1,n_2}\int\dx \bx_1 \dx \bx_2 A(\bx,\bx_1,n,n_1)V(\bx_1,\bx_2,n_1,n_2)B(\bx_2,\bx',n_2,n').
\end{equation}
In this notation,  the Dyson equation for the Green's function of the time-modulated problem becomes
\begin{equation}\label{eq:dyson}
	G =  G_0 +  G_0V G,
\end{equation}
where $ G_0$ is the free-space Green's function for the Helmholtz equation,
\begin{equation}
	G_0(\bx,\bx',n,n') = \delta_{nn'}\frac{e^{\iu k_{n'}|\bx-\bx'|}}{4\pi|\bx-\bx'|}.
\end{equation}
Equation \eqref{eq:dyson} is a direct generalization of the Dyson equation for the Helmholtz equation without time-modulation (see, for example, \cite{carminati2021principles}). This also defines a Born series for $ G$:
\begin{equation}
	G =  G_0 +  G_0V G_0 +  G_0V G_0V G_0 + \cdots ,
\end{equation}
which may be rephrased in terms of the $T$-matrix as $ G =  G_0 +  G_0T G_0$, where $T$ is given by the iterated series 
\begin{equation}
	T = V + V  G_0 V + V  G_0 V  G_0 V  + \cdots .
\end{equation}
We consider a point-scattering model for the tip, whose $T$-matrix given by 
\begin{equation}
	T (\bx_1,\bx_2,n_1,n_2) = \delta(\bx_1-\bx_0)\delta(\bx_2-\bx_0)\alpha_{n_1-n_2}k_{n_1}^2,
\end{equation}
where $\alpha_{-n} = \overline{\alpha_{n}}$ corresponding to a time-dependent, nonabsorbing point scatterer at $\bx=\bx_0$. It follows that the combined $T$-matrix can be written as an iterated series over $S$,
\begin{equation}\label{eq:Texp}
	T = S + T  + S T  + T S + T ST  + \cdots .
\end{equation}
In terms of the $T$-matrix,  the scattered field $\svsc$ can now be written 
\begin{align}
	\svsc &=  G_0T \svin \\
	&= \sv^{\textrm{S}}+\sv^{\textrm{T}}+\sv^{\textrm{ST}}+\sv^{\textrm{TS}}+\sv^{\textrm{TST}} + \cdots,
\end{align}
where each term originates from corresponding term of the $T$-matrix. In particular, using the far-field expansion $G^{k_n}(\bx-\bx')\sim G^{k_n}(\bx)e^{-\iu k_n\hat{\bx}\cdot \bx'}$, along with the definition \eqref{eq:vin} of the incident field, we have a representation of the far-field amplitude $A_{ij}(\hat{\bx},\hat{\bz})$ in terms of the $T$-matrix as follows,
\begin{equation}\label{eq:AT}
	A_{ij}(\hat{\bx},\hat{\bz}) = \frac{1}{4\pi}\int \dx\bx_1 \dx\bx_2e^{-\iu k_i \hat{\bx}\cdot \bx_1}T(\bx_1,\bx_2,i,j)e^{\iu k_j \hat{\bz}\cdot \bx_2}.
\end{equation}
Based on the decomposition \eqref{eq:Texp}, we may split $A$ into corresponding terms $A^\textrm{S}$, $A^\textrm{T}$, $A^\textrm{ST}$, $A^\textrm{TS}$, and $A^\textrm{TST}$.

Again, we assume that the sample is non-absorbing, so that $s(\bx,t)$ is real-valued and $s_{-i}(\bx) = \overline{s_i(\bx)}$, and seek to reconstruct the off-diagonal entries $s_i$ for $i\neq 0$. As in Section \ref{sec:born}, we take an incident field of two plane waves,
\begin{equation}
	\vin_i(\bx) = c_1\delta_{ij_1} e^{\iu k_{j_1} \hat{\bz}_1\cdot \bx} + c_2\delta_{ij_2} e^{\iu k_{j_2}\hat{\bz}_2\cdot \bx},
\end{equation} 
and let $\bk_n = k_{j_n} \hat{\bz}_n$. Throughout, we assume that $j_1\neq j_2$. This again gives $\Pext(c_1,c_2) =\bc^* M \bc$, where the entries of $M$ are now given by 
\begin{equation}\label{eq:Pmn}
	M_{mn} = \frac{4\pi}{2\iu}\left( k_{j_m}A_{j_m j_n}(\hat{\bz}_m,\hat{\bz}_n) - k_{j_n}\overline{A_{j_n j_m}}(\hat{\bz}_n,\hat{\bz}_m)\right),
\end{equation}
for $m,n =1,2$. To recover the off-diagonal entries of $M$, we need four measurements of $\Pext$. We define the data function as 
\begin{equation}\label{eq:D_NSOM}
	D(\hat{\bz}_1,\hat{\bz}_2) = \frac{1}{4}\Bigl(\Pext(1,1) - \Pext(1,-1) - \iu \left(\Pext(1,\iu) - \Pext(1,-\iu)\right)\Bigr),
\end{equation}
which, based on \eqref{eq:Pmn}, gives $D(\hat{\bz}_1,\hat{\bz}_2) = M_{12}$. 

For the reconstruction step, we seek to recover $\eta_{j_1-j_2}$ based on \eqref{eq:Pmn} for $M_{12}$. While \eqref{eq:Texp} is non-perturbative in $T $, we take a weak-scattering approach where we neglect $TST$ and higher terms. As we shall see, the terms $ST$ and $TS$, which is of lower order than $TST$, will be sufficient for reconstruction of the time-varying part.

The terms $S$ and $T $, corresponding to the sample and the tip without interaction, respectively, can be considered a known background contribution. By subtracting a previously measured field, these terms  may be omitted, and we consider $A = A^\textrm{ST} + A^\textrm{TS}$. In this setting, we obtain from \eqref{eq:AT}
\begin{multline}\label{eq:A}
	A_{mn}(\hat{\bx},\hat{\bz}) = k_{m}^2\frac{e^{\iu k_n\hat{\bz}\cdot \bx_0}}{4\pi} \sum_{n_1}\int \dx \bx_1  e^{-\iu k_m \hat{\bx}\cdot \bx_1 } s_{m-n_1}(\bx_1 ) k_{n_1}^2 G^{k_{n_1}}(\bx_1 -\bx_0)\alpha_{n_1-n} \\
	+ k_{m}^2\frac{e^{-\iu k_m\hat{\bx}\cdot \bx_0}}{4\pi} \sum_{n_1}\int \dx \bx_1  e^{\iu k_n \hat{\bz}\cdot \bx_1 } s_{n_1-n}(\bx_1 ) k_{n_1}^2 G^{k_{m}}(\bx_1 -\bx_0)\alpha_{m-n_1}.
\end{multline}
Up to this point, the calculations hold for a tip which itself may be time-modulated. We proceed by simplifying this expression in the case $\alpha_n =\delta_{n0}\alpha_0$, corresponding to a tip which is static on the time-scale of the sample modulation. Using \eqref{eq:Pmn}, this gives
\begin{multline}\label{eq:M12}
	M_{12}(\bx_0) =\alpha_0  \Bigg(e^{\iu \bk_2\cdot \bx_0} \int \dx \bx  e^{-\iu \bk_1\cdot \bx } s_{j_1-j_2}(\bx ) k_{j_2}^2A(\bx -\bx_0)  \\
	+ e^{-\iu \bk_1\cdot \bx_0} \int \dx \bx  e^{\iu \bk_2\cdot \bx } s_{j_1-j_2}(\bx ) k_{j_1}^2B(\bx -\bx_0)\Bigg), 
\end{multline}
where 
\begin{equation}
	A(\bx) = \frac{1}{2\iu}\left(k_{j_1}^3G^{k_{j_2}}(\bx)- k_{j_2}^3(G^{k_{j_2}})^*(\bx)\right), \qquad B(\bx) = \frac{1}{2\iu}\left(k_{j_1}^3G^{k_{j_1}}(\bx)-k_{j_2}^3(G^{k_{j_1}})^*(\bx)\right).
\end{equation}
For illustration, we consider the case of a planar, two-dimensional sample, where 
\begin{equation}
	s_j(\bx) = \delta(z)s_j(x,y).
\end{equation}
We let $P_{xy}$ denote the projection onto the $xy$-plane, and define 
\begin{equation}
	\bx_\parallel = P_{xy}[\bx],\quad \bx_{0\parallel} = P_{xy}[\bx_0], \quad \bk_{n\parallel} = P_{xy}[\bk_n].
\end{equation}
We will simplify \eqref{eq:M12} using Weyl's identity, which is a decomposition of the Green's function into plane waves and evanescent waves, \cite{wolf1969three,weyl1919ausbreitung}
\begin{equation}
	G^k(\bx) = \frac{\iu}{8\pi^2}\int \dx \bq  \: e^{\iu \bq \cdot \bx_\parallel}\frac{e^{\iu k_z(\bq,) |z|}}{k_z(k,\bq)},
\end{equation}
where $k_z(\bq,k)$ is the function
\begin{equation}
	k_z(\bq,k) = \begin{cases} \sqrt{k^2 - |\bq|^2}, \quad & k \geq |\bq|, \\
		\iu \sqrt{|\bq|^2 - k^2}, & k < |\bq|.\end{cases}
\end{equation}
This gives, after simplifications,
\begin{equation}\label{eq:Mtilde}
		\widetilde{M}(\bq  - \bk_{1\parallel}+ \bk_{2\parallel}) = F(\bq) \widetilde{s}_{j_1-j_2}(\bq),
\end{equation}
where $\widetilde{M}$ and $\widetilde{s}$ denote the Fourier transforms of $M_{12}$ and $s$, respectively, in the $xy$-plane, and where
\begin{multline}
	F(\bq) = \frac{\alpha_0}{4}\Bigg[\left( \frac{k_{j_2}^2k_{j_1}^3}{k_{z,2}(\bq)}e^{\iu k_{z,2}(\bq)|z_0|} +\frac{k_{j_2}^5}{\overline{k_{z,2}}(\bq)}e^{-\iu \overline{k_{z,2}}(\bq)|z_0|}\right)e^{\iu (\bk_2)_zz_0}\\
	+\left( \frac{k_{j_1}^5}{k_{z,1}(\bq)}e^{\iu k_{z,1}(\bq)|z_0|} +\frac{k_{j_1}^2k_{j_2}^3}{\overline{k_{z,1}(\bq)}}e^{-\iu \overline{k_{z,1}}(\bq)|z_0|}\right)e^{-\iu (\bk_1)_zz_0}\Bigg],
\end{multline}
for $k_{z,1}(\bq) = k_z(\bk_{2\parallel} + \bq,k_{j_1})$ and $k_{z,2}(\bq) = k_z(\bk_{1\parallel} - \bq,k_{j_2})$.

As long as $F(\bq)$ is nonzero, equation \eqref{eq:Mtilde} directly provides the solution $\widetilde{s}_{j_1-j_2}$ to the inverse problem. We note that $F$ is evanescent for large $|\bq|$, when both $k_{z,1}$ and $k_{z,2}$ become imaginary. This produces an upper band-limit of the reconstructed image. However, for small $z_0$ (for a probing tip close to the sample), we obtain subwavelength resolution in the same manner as for the inverse problem without time-modulation \cite{sun2006near,sun2007strong}. Moreover, for small $|\bq|$, we note that at least one value of $k_{z,1}$ and $k_{z,2}$ will be real, and the lower band limit of Section \ref{sec:born} is no longer affecting the reconstructed image.

For $j_1 = j_2$, that is for imaging of the constant component $s_0$ of $s$, we find that $F(\bq)$ vanishes for $|\bq|$ such that $k_{z,1}$ and $k_{z,2}$ are both imaginary. In this case, it is necessary to use a strong scattering tip and consider the term originating from $TST$ in \eqref{eq:Texp}. Neglecting lower order terms, we now have
\begin{equation}\label{eq:MtildeTST}
	\widetilde{M}^\mathrm{TST}(\bq - \bk_{1\parallel}+ \bk_{2\parallel}) = F^\mathrm{TST}(\bq) \widetilde{s}_{j_1-j_2}(\bq),
\end{equation}
where 
\begin{multline}
	F^\mathrm{TST}(\bq) = \frac{\alpha_0^2}{4}e^{\iu (\bk_2-\bk_1)_zz_0}\int \dx \bq' \Bigg( \frac{k_{j_1}^5k_{j_2}^2}{k_{z}(\bq-\bq',k_{j_1})k_{z}(\bq',k_{j_2})}e^{\iu \bigl(k_{z}(\bq-\bq',k_{j_1})+k_{z}(\bq',k_{j_2})\bigr)|z_0|} \\
	+ \frac{k_{j_1}^2k_{j_2}^5}{k_{z}(\bq-\bq',k_{j_1})^*k_{z}(\bq',k_{j_2})^*}e^{-\iu \bigl(k_{z}(\bq-\bq',k_{j_1})^*+k_{z}(\bq',k_{j_2})^*\bigr)|z_0|}\Bigg).
\end{multline}
Again, we have that $F^\mathrm{TST}(\bq)$ is evanescent for large $|\bq|$, but subwavelength resolution can be achieved using small $|z_0|$.

We conclude this section by commenting on the case of three-dimensional imaging. We suppose that  $j_1 \neq j_2$, where the terms $TS$ and $ST$ are then sufficient for the purposes of reconstruction. We follow the approach of \cite{govyadinov2009phaseless}, where the inverse problem is now phrased in terms of a one-dimensional integral equation in the $z$-coordinate:
\begin{equation}
	\widetilde{M}(\bq - \bk_{1\parallel}+ \bk_{2\parallel},z_0) = \int_V \dx z F(\bq,z,z_0) \widetilde{s}_{j_1-j_2}(\bq,z),
\end{equation}
where (as before) $\widetilde{M}$ and $\widetilde{s}$ denote the Fourier transforms of $M_{12}$ and $s$, respectively, in the $xy$-plane, and where the kernel $F$ is given by
\begin{multline}
	F(\bq,z,z_0) = \frac{\alpha_0}{4}\Bigg[\left( \frac{k_{j_2}^2k_{j_1}^3}{k_{z,2}(\bq)}e^{\iu k_{z,2}(\bq)|z-z_0|} +\frac{k_{j_2}^5}{\overline{k_{z,2}}(\bq)}e^{-\iu \overline{k_{z,2}}(\bq)|z-z_0|}\right)e^{\iu (\bk_2)_zz_0-\iu (\bk_1)_zz}\\
	+\left( \frac{k_{j_1}^5}{k_{z,1}(\bq)}e^{\iu k_{z,1}(\bq)|z-z_0|} +\frac{k_{j_1}^2k_{j_2}^3}{\overline{k_{z,1}(\bq)}}e^{-\iu \overline{k_{z,1}}(\bq)|z-z_0|}\right)e^{\iu(\bk_2)_zz-\iu (\bk_1)_zz_0}\Bigg].
\end{multline}
Here $V\subset \mathbb{R}$ denotes the projection of the support of $s_{j_1-j_2}(\bx)$ onto the $z$-axis and we restrict to $z_0\notin V$, so that the tip does not overlap the sample. As in \cite{govyadinov2009phaseless}, this problem is ill-posed, and regularization techniques are necessary. The details of this problem are left for future studies.

\section{Concluding remarks}\label{sec:concl}
We have studied the energy balance of a time-modulated material and derived an optical theorem for such systems. Our approach accounts for the broken energy conservation which arises due to frequency conversion, and we have shown that a conserved quantity is obtained by redefining the weights of the power of each frequency harmonic. We have seen that the same system might be subject to energy gain, energy absorption or energy loss, depending on the ratio of these harmonics present in the total field. We expect that these results will be significant for developing a general theory for energy-preserving spacetime metamaterials, with the goal to overcome instability and high energy cost associated to these systems. We have also shown how the optical theorem may be used as starting-point for image reconstruction based on power measurements. For the linear inverse scattering problem, the reconstructed image is subject to both upper and lower band-limits. We have shown that this can be mitigated using near-field microscopy, where the time-dependent sample is scanned by a probing tip in the near-field and the interaction between the tip and the sample is measured. 

\section{Data availability}
The data that support the findings of this article are openly available \cite{code}.

\bibliography{refs}

\begin{thebibliography}{28}%
\makeatletter
\providecommand \@ifxundefined [1]{%
 \@ifx{#1\undefined}
}%
\providecommand \@ifnum [1]{%
 \ifnum #1\expandafter \@firstoftwo
 \else \expandafter \@secondoftwo
 \fi
}%
\providecommand \@ifx [1]{%
 \ifx #1\expandafter \@firstoftwo
 \else \expandafter \@secondoftwo
 \fi
}%
\providecommand \natexlab [1]{#1}%
\providecommand \enquote  [1]{``#1''}%
\providecommand \bibnamefont  [1]{#1}%
\providecommand \bibfnamefont [1]{#1}%
\providecommand \citenamefont [1]{#1}%
\providecommand \href@noop [0]{\@secondoftwo}%
\providecommand \href [0]{\begingroup \@sanitize@url \@href}%
\providecommand \@href[1]{\@@startlink{#1}\@@href}%
\providecommand \@@href[1]{\endgroup#1\@@endlink}%
\providecommand \@sanitize@url [0]{\catcode `\\12\catcode `\$12\catcode
  `\&12\catcode `\#12\catcode `\^12\catcode `\_12\catcode `\%12\relax}%
\providecommand \@@startlink[1]{}%
\providecommand \@@endlink[0]{}%
\providecommand \url  [0]{\begingroup\@sanitize@url \@url }%
\providecommand \@url [1]{\endgroup\@href {#1}{\urlprefix }}%
\providecommand \urlprefix  [0]{URL }%
\providecommand \Eprint [0]{\href }%
\providecommand \doibase [0]{https://doi.org/}%
\providecommand \selectlanguage [0]{\@gobble}%
\providecommand \bibinfo  [0]{\@secondoftwo}%
\providecommand \bibfield  [0]{\@secondoftwo}%
\providecommand \translation [1]{[#1]}%
\providecommand \BibitemOpen [0]{}%
\providecommand \bibitemStop [0]{}%
\providecommand \bibitemNoStop [0]{.\EOS\space}%
\providecommand \EOS [0]{\spacefactor3000\relax}%
\providecommand \BibitemShut  [1]{\csname bibitem#1\endcsname}%
\let\auto@bib@innerbib\@empty
\bibitem [{\citenamefont {Holberg}\ and\ \citenamefont
  {Kunz}(1966)}]{holberg1966parametric}%
  \BibitemOpen
  \bibfield  {author} {\bibinfo {author} {\bibfnamefont {D.}~\bibnamefont
  {Holberg}}\ and\ \bibinfo {author} {\bibfnamefont {K.}~\bibnamefont {Kunz}},\
  }\bibfield  {title} {\bibinfo {title} {Parametric properties of fields in a
  slab of time-varying permittivity},\ }\href@noop {} {\bibfield  {journal}
  {\bibinfo  {journal} {IEEE Transactions on Antennas and Propagation}\
  }\textbf {\bibinfo {volume} {14}},\ \bibinfo {pages} {183} (\bibinfo {year}
  {1966})}\BibitemShut {NoStop}%
\bibitem [{\citenamefont {Koutserimpas}\ \emph {et~al.}(2018)\citenamefont
  {Koutserimpas}, \citenamefont {Al{\`u}},\ and\ \citenamefont
  {Fleury}}]{koutserimpas2018parametric}%
  \BibitemOpen
  \bibfield  {author} {\bibinfo {author} {\bibfnamefont {T.~T.}\ \bibnamefont
  {Koutserimpas}}, \bibinfo {author} {\bibfnamefont {A.}~\bibnamefont
  {Al{\`u}}},\ and\ \bibinfo {author} {\bibfnamefont {R.}~\bibnamefont
  {Fleury}},\ }\bibfield  {title} {\bibinfo {title} {Parametric amplification
  and bidirectional invisibility in pt-symmetric time-floquet systems},\
  }\href@noop {} {\bibfield  {journal} {\bibinfo  {journal} {Physical Review
  A}\ }\textbf {\bibinfo {volume} {97}},\ \bibinfo {pages} {013839} (\bibinfo
  {year} {2018})}\BibitemShut {NoStop}%
\bibitem [{\citenamefont {Zhang}\ \emph {et~al.}(2024)\citenamefont {Zhang},
  \citenamefont {Donaldson},\ and\ \citenamefont
  {Agrawal}}]{zhang2024conservation}%
  \BibitemOpen
  \bibfield  {author} {\bibinfo {author} {\bibfnamefont {J.}~\bibnamefont
  {Zhang}}, \bibinfo {author} {\bibfnamefont {W.}~\bibnamefont {Donaldson}},\
  and\ \bibinfo {author} {\bibfnamefont {G.~P.}\ \bibnamefont {Agrawal}},\
  }\bibfield  {title} {\bibinfo {title} {Conservation law for electromagnetic
  fields in a space-time-varying medium and its implications},\ }\href@noop {}
  {\bibfield  {journal} {\bibinfo  {journal} {Physical Review A}\ }\textbf
  {\bibinfo {volume} {110}},\ \bibinfo {pages} {043526} (\bibinfo {year}
  {2024})}\BibitemShut {NoStop}%
\bibitem [{\citenamefont {Liberal}\ \emph {et~al.}(2024)\citenamefont
  {Liberal}, \citenamefont {Ganfornina-Andrades},\ and\ \citenamefont
  {V{\'a}zquez-Lozano}}]{liberal2024spatiotemporal}%
  \BibitemOpen
  \bibfield  {author} {\bibinfo {author} {\bibfnamefont {I.}~\bibnamefont
  {Liberal}}, \bibinfo {author} {\bibfnamefont {A.}~\bibnamefont
  {Ganfornina-Andrades}},\ and\ \bibinfo {author} {\bibfnamefont {J.~E.}\
  \bibnamefont {V{\'a}zquez-Lozano}},\ }\bibfield  {title} {\bibinfo {title}
  {Spatiotemporal symmetries and energy-momentum conservation in uniform
  spacetime metamaterials},\ }\href@noop {} {\bibfield  {journal} {\bibinfo
  {journal} {ACS photonics}\ }\textbf {\bibinfo {volume} {11}},\ \bibinfo
  {pages} {5273} (\bibinfo {year} {2024})}\BibitemShut {NoStop}%
\bibitem [{\citenamefont {Zhang}\ \emph {et~al.}(2023)\citenamefont {Zhang},
  \citenamefont {Donaldson},\ and\ \citenamefont
  {Agrawal}}]{zhang2023generalized}%
  \BibitemOpen
  \bibfield  {author} {\bibinfo {author} {\bibfnamefont {J.}~\bibnamefont
  {Zhang}}, \bibinfo {author} {\bibfnamefont {W.~R.}\ \bibnamefont
  {Donaldson}},\ and\ \bibinfo {author} {\bibfnamefont {G.~P.}\ \bibnamefont
  {Agrawal}},\ }\bibfield  {title} {\bibinfo {title} {Generalized energy
  conservation relation in a space-time varying medium},\ }in\ \href@noop {}
  {\emph {\bibinfo {booktitle} {2023 Conference on Lasers and Electro-Optics
  (CLEO)}}}\ (\bibinfo {organization} {IEEE},\ \bibinfo {year} {2023})\ pp.\
  \bibinfo {pages} {1--2}\BibitemShut {NoStop}%
\bibitem [{\citenamefont {Hirosawa}\ and\ \citenamefont
  {Wirth}(2009)}]{hirosawa2009generalised}%
  \BibitemOpen
  \bibfield  {author} {\bibinfo {author} {\bibfnamefont {F.}~\bibnamefont
  {Hirosawa}}\ and\ \bibinfo {author} {\bibfnamefont {J.}~\bibnamefont
  {Wirth}},\ }\bibfield  {title} {\bibinfo {title} {Generalised energy
  conservation law for wave equations with variable propagation speed},\
  }\href@noop {} {\bibfield  {journal} {\bibinfo  {journal} {Journal of
  mathematical analysis and applications}\ }\textbf {\bibinfo {volume} {358}},\
  \bibinfo {pages} {56} (\bibinfo {year} {2009})}\BibitemShut {NoStop}%
\bibitem [{\citenamefont {Ebert}\ \emph {et~al.}(2015)\citenamefont {Ebert},
  \citenamefont {Fitriana},\ and\ \citenamefont {Hirosawa}}]{ebert2015energy}%
  \BibitemOpen
  \bibfield  {author} {\bibinfo {author} {\bibfnamefont {M.~R.}\ \bibnamefont
  {Ebert}}, \bibinfo {author} {\bibfnamefont {L.}~\bibnamefont {Fitriana}},\
  and\ \bibinfo {author} {\bibfnamefont {F.}~\bibnamefont {Hirosawa}},\
  }\bibfield  {title} {\bibinfo {title} {On the energy estimates of the wave
  equation with time dependent propagation speed asymptotically monotone
  functions},\ }\href@noop {} {\bibfield  {journal} {\bibinfo  {journal}
  {Journal of Mathematical Analysis and Applications}\ }\textbf {\bibinfo
  {volume} {432}},\ \bibinfo {pages} {654} (\bibinfo {year}
  {2015})}\BibitemShut {NoStop}%
\bibitem [{\citenamefont {Cullen}(1958)}]{cullen1958travelling}%
  \BibitemOpen
  \bibfield  {author} {\bibinfo {author} {\bibfnamefont {A.}~\bibnamefont
  {Cullen}},\ }\bibfield  {title} {\bibinfo {title} {A travelling-wave
  parametric amplifier},\ }\href@noop {} {\bibfield  {journal} {\bibinfo
  {journal} {Nature}\ }\textbf {\bibinfo {volume} {181}},\ \bibinfo {pages}
  {332} (\bibinfo {year} {1958})}\BibitemShut {NoStop}%
\bibitem [{\citenamefont {Raiford}(1974)}]{raiford1974degenerate}%
  \BibitemOpen
  \bibfield  {author} {\bibinfo {author} {\bibfnamefont {M.}~\bibnamefont
  {Raiford}},\ }\bibfield  {title} {\bibinfo {title} {Degenerate parametric
  amplification with time-dependent pump amplitude and phase},\ }\href@noop {}
  {\bibfield  {journal} {\bibinfo  {journal} {Phys. Rev. A}\ }\textbf {\bibinfo
  {volume} {9}},\ \bibinfo {pages} {2060} (\bibinfo {year} {1974})}\BibitemShut
  {NoStop}%
\bibitem [{\citenamefont {Koufidis}\ \emph {et~al.}(2024)\citenamefont
  {Koufidis}, \citenamefont {Koutserimpas}, \citenamefont {Monticone},\ and\
  \citenamefont {McCall}}]{koufidis2024enhanced}%
  \BibitemOpen
  \bibfield  {author} {\bibinfo {author} {\bibfnamefont {S.~F.}\ \bibnamefont
  {Koufidis}}, \bibinfo {author} {\bibfnamefont {T.~T.}\ \bibnamefont
  {Koutserimpas}}, \bibinfo {author} {\bibfnamefont {F.}~\bibnamefont
  {Monticone}},\ and\ \bibinfo {author} {\bibfnamefont {M.~W.}\ \bibnamefont
  {McCall}},\ }\bibfield  {title} {\bibinfo {title} {Enhanced scattering from
  an almost-periodic optical temporal slab},\ }\href@noop {} {\bibfield
  {journal} {\bibinfo  {journal} {Physical Review A}\ }\textbf {\bibinfo
  {volume} {110}},\ \bibinfo {pages} {L041501} (\bibinfo {year}
  {2024})}\BibitemShut {NoStop}%
\bibitem [{\citenamefont {Koutserimpas}(2022)}]{koutserimpas2022parametric}%
  \BibitemOpen
  \bibfield  {author} {\bibinfo {author} {\bibfnamefont {T.~T.}\ \bibnamefont
  {Koutserimpas}},\ }\bibfield  {title} {\bibinfo {title} {Parametric
  amplification interactions in time-periodic media: coupled waves theory},\
  }\href@noop {} {\bibfield  {journal} {\bibinfo  {journal} {JOSA B}\ }\textbf
  {\bibinfo {volume} {39}},\ \bibinfo {pages} {481} (\bibinfo {year}
  {2022})}\BibitemShut {NoStop}%
\bibitem [{\citenamefont {Demkowicz}\ \emph {et~al.}(2024)\citenamefont
  {Demkowicz}, \citenamefont {Melenk}, \citenamefont {Badger},\ and\
  \citenamefont {Henneking}}]{demkowicz2024stability}%
  \BibitemOpen
  \bibfield  {author} {\bibinfo {author} {\bibfnamefont {L.}~\bibnamefont
  {Demkowicz}}, \bibinfo {author} {\bibfnamefont {J.~M.}\ \bibnamefont
  {Melenk}}, \bibinfo {author} {\bibfnamefont {J.}~\bibnamefont {Badger}},\
  and\ \bibinfo {author} {\bibfnamefont {S.}~\bibnamefont {Henneking}},\
  }\bibfield  {title} {\bibinfo {title} {Stability analysis for electromagnetic
  waveguides. part 2: non-homogeneous waveguides},\ }\href@noop {} {\bibfield
  {journal} {\bibinfo  {journal} {Advances in Computational Mathematics}\
  }\textbf {\bibinfo {volume} {50}},\ \bibinfo {pages} {35} (\bibinfo {year}
  {2024})}\BibitemShut {NoStop}%
\bibitem [{\citenamefont {Yerezhep}\ and\ \citenamefont
  {Valagiannopoulos}(2021)}]{yerezhep2021approximate}%
  \BibitemOpen
  \bibfield  {author} {\bibinfo {author} {\bibfnamefont {B.}~\bibnamefont
  {Yerezhep}}\ and\ \bibinfo {author} {\bibfnamefont {C.}~\bibnamefont
  {Valagiannopoulos}},\ }\bibfield  {title} {\bibinfo {title} {Approximate
  stability dynamics of concentric cylindrical metasurfaces},\ }\href@noop {}
  {\bibfield  {journal} {\bibinfo  {journal} {IEEE Transactions on Antennas and
  Propagation}\ }\textbf {\bibinfo {volume} {69}},\ \bibinfo {pages} {5716}
  (\bibinfo {year} {2021})}\BibitemShut {NoStop}%
\bibitem [{\citenamefont {Hiltunen}\ and\ \citenamefont
  {Davies}(2024)}]{hiltunen2024coupled}%
  \BibitemOpen
  \bibfield  {author} {\bibinfo {author} {\bibfnamefont {E.~O.}\ \bibnamefont
  {Hiltunen}}\ and\ \bibinfo {author} {\bibfnamefont {B.}~\bibnamefont
  {Davies}},\ }\bibfield  {title} {\bibinfo {title} {Coupled harmonics due to
  time-modulated point scatterers},\ }\href@noop {} {\bibfield  {journal}
  {\bibinfo  {journal} {Physical Review B}\ }\textbf {\bibinfo {volume}
  {110}},\ \bibinfo {pages} {184102} (\bibinfo {year} {2024})}\BibitemShut
  {NoStop}%
\bibitem [{\citenamefont {Ammari}\ \emph {et~al.}(2023)\citenamefont {Ammari},
  \citenamefont {Cao}, \citenamefont {Hiltunen},\ and\ \citenamefont
  {Rueff}}]{ammari2023transmission}%
  \BibitemOpen
  \bibfield  {author} {\bibinfo {author} {\bibfnamefont {H.}~\bibnamefont
  {Ammari}}, \bibinfo {author} {\bibfnamefont {J.}~\bibnamefont {Cao}},
  \bibinfo {author} {\bibfnamefont {E.~O.}\ \bibnamefont {Hiltunen}},\ and\
  \bibinfo {author} {\bibfnamefont {L.}~\bibnamefont {Rueff}},\ }\bibfield
  {title} {\bibinfo {title} {Transmission properties of time-dependent
  one-dimensional metamaterials},\ }\href@noop {} {\bibfield  {journal}
  {\bibinfo  {journal} {Journal of Mathematical Physics}\ }\textbf {\bibinfo
  {volume} {64}} (\bibinfo {year} {2023})}\BibitemShut {NoStop}%
\bibitem [{\citenamefont {Deshmukh}\ and\ \citenamefont
  {Milton}(2022)}]{deshmukh2022energy}%
  \BibitemOpen
  \bibfield  {author} {\bibinfo {author} {\bibfnamefont {K.~J.}\ \bibnamefont
  {Deshmukh}}\ and\ \bibinfo {author} {\bibfnamefont {G.~W.}\ \bibnamefont
  {Milton}},\ }\bibfield  {title} {\bibinfo {title} {An energy conserving
  mechanism for temporal metasurfaces},\ }\href@noop {} {\bibfield  {journal}
  {\bibinfo  {journal} {Applied Physics Letters}\ }\textbf {\bibinfo {volume}
  {121}} (\bibinfo {year} {2022})}\BibitemShut {NoStop}%
\bibitem [{\citenamefont {Horsley}\ and\ \citenamefont
  {Pendry}(2023)}]{horsley2023quantum}%
  \BibitemOpen
  \bibfield  {author} {\bibinfo {author} {\bibfnamefont {S.~A.}\ \bibnamefont
  {Horsley}}\ and\ \bibinfo {author} {\bibfnamefont {J.~B.}\ \bibnamefont
  {Pendry}},\ }\bibfield  {title} {\bibinfo {title} {Quantum electrodynamics of
  time-varying gratings},\ }\href@noop {} {\bibfield  {journal} {\bibinfo
  {journal} {Proceedings of the National Academy of Sciences}\ }\textbf
  {\bibinfo {volume} {120}},\ \bibinfo {pages} {e2302652120} (\bibinfo {year}
  {2023})}\BibitemShut {NoStop}%
\bibitem [{\citenamefont {Carney}\ \emph {et~al.}(2001)\citenamefont {Carney},
  \citenamefont {Markel},\ and\ \citenamefont {Schotland}}]{carney2001near}%
  \BibitemOpen
  \bibfield  {author} {\bibinfo {author} {\bibfnamefont {P.~S.}\ \bibnamefont
  {Carney}}, \bibinfo {author} {\bibfnamefont {V.~A.}\ \bibnamefont {Markel}},\
  and\ \bibinfo {author} {\bibfnamefont {J.~C.}\ \bibnamefont {Schotland}},\
  }\bibfield  {title} {\bibinfo {title} {Near-field tomography without phase
  retrieval},\ }\href@noop {} {\bibfield  {journal} {\bibinfo  {journal}
  {Physical review letters}\ }\textbf {\bibinfo {volume} {86}},\ \bibinfo
  {pages} {5874} (\bibinfo {year} {2001})}\BibitemShut {NoStop}%
\bibitem [{\citenamefont {Govyadinov}\ \emph {et~al.}(2009)\citenamefont
  {Govyadinov}, \citenamefont {Panasyuk},\ and\ \citenamefont
  {Schotland}}]{govyadinov2009phaseless}%
  \BibitemOpen
  \bibfield  {author} {\bibinfo {author} {\bibfnamefont {A.~A.}\ \bibnamefont
  {Govyadinov}}, \bibinfo {author} {\bibfnamefont {G.~Y.}\ \bibnamefont
  {Panasyuk}},\ and\ \bibinfo {author} {\bibfnamefont {J.~C.}\ \bibnamefont
  {Schotland}},\ }\bibfield  {title} {\bibinfo {title} {Phaseless
  three-dimensional optical nanoimaging},\ }\href@noop {} {\bibfield  {journal}
  {\bibinfo  {journal} {Physical review letters}\ }\textbf {\bibinfo {volume}
  {103}},\ \bibinfo {pages} {213901} (\bibinfo {year} {2009})}\BibitemShut
  {NoStop}%
\bibitem [{\citenamefont {Carney}(1999)}]{carney1999optical}%
  \BibitemOpen
  \bibfield  {author} {\bibinfo {author} {\bibfnamefont {P.~S.}\ \bibnamefont
  {Carney}},\ }\bibfield  {title} {\bibinfo {title} {The optical cross-section
  theorem with incident fields containing evanescent components},\ }\href@noop
  {} {\bibfield  {journal} {\bibinfo  {journal} {journal of modern optics}\
  }\textbf {\bibinfo {volume} {46}},\ \bibinfo {pages} {891} (\bibinfo {year}
  {1999})}\BibitemShut {NoStop}%
\bibitem [{\citenamefont {Carminati}\ and\ \citenamefont
  {Schotland}(2021)}]{carminati2021principles}%
  \BibitemOpen
  \bibfield  {author} {\bibinfo {author} {\bibfnamefont {R.}~\bibnamefont
  {Carminati}}\ and\ \bibinfo {author} {\bibfnamefont {J.~C.}\ \bibnamefont
  {Schotland}},\ }\href@noop {} {\emph {\bibinfo {title} {Principles of
  scattering and transport of light}}}\ (\bibinfo  {publisher} {Cambridge
  University Press},\ \bibinfo {year} {2021})\BibitemShut {NoStop}%
\bibitem [{\citenamefont {Born}\ and\ \citenamefont
  {Wolf}(2013)}]{born2013principles}%
  \BibitemOpen
  \bibfield  {author} {\bibinfo {author} {\bibfnamefont {M.}~\bibnamefont
  {Born}}\ and\ \bibinfo {author} {\bibfnamefont {E.}~\bibnamefont {Wolf}},\
  }\href@noop {} {\emph {\bibinfo {title} {Principles of optics:
  electromagnetic theory of propagation, interference and diffraction of
  light}}}\ (\bibinfo  {publisher} {Elsevier},\ \bibinfo {year}
  {2013})\BibitemShut {NoStop}%
\bibitem [{\citenamefont {Fleury}\ \emph {et~al.}(2016)\citenamefont {Fleury},
  \citenamefont {Khanikaev},\ and\ \citenamefont {Alu}}]{fleury2016floquet}%
  \BibitemOpen
  \bibfield  {author} {\bibinfo {author} {\bibfnamefont {R.}~\bibnamefont
  {Fleury}}, \bibinfo {author} {\bibfnamefont {A.~B.}\ \bibnamefont
  {Khanikaev}},\ and\ \bibinfo {author} {\bibfnamefont {A.}~\bibnamefont
  {Alu}},\ }\bibfield  {title} {\bibinfo {title} {Floquet topological
  insulators for sound},\ }\href@noop {} {\bibfield  {journal} {\bibinfo
  {journal} {Nature communications}\ }\textbf {\bibinfo {volume} {7}},\
  \bibinfo {pages} {11744} (\bibinfo {year} {2016})}\BibitemShut {NoStop}%
\bibitem [{\citenamefont {Sun}\ \emph {et~al.}(2006)\citenamefont {Sun},
  \citenamefont {Carney},\ and\ \citenamefont {Schotland}}]{sun2006near}%
  \BibitemOpen
  \bibfield  {author} {\bibinfo {author} {\bibfnamefont {J.}~\bibnamefont
  {Sun}}, \bibinfo {author} {\bibfnamefont {P.~S.}\ \bibnamefont {Carney}},\
  and\ \bibinfo {author} {\bibfnamefont {J.~C.}\ \bibnamefont {Schotland}},\
  }\bibfield  {title} {\bibinfo {title} {Near-field scanning optical
  tomography: a nondestructive method for three-dimensional nanoscale
  imaging},\ }\href@noop {} {\bibfield  {journal} {\bibinfo  {journal} {IEEE
  Journal of selected topics in quantum electronics}\ }\textbf {\bibinfo
  {volume} {12}},\ \bibinfo {pages} {1072} (\bibinfo {year}
  {2006})}\BibitemShut {NoStop}%
\bibitem [{\citenamefont {Sun}\ \emph {et~al.}(2007)\citenamefont {Sun},
  \citenamefont {Carney},\ and\ \citenamefont {Schotland}}]{sun2007strong}%
  \BibitemOpen
  \bibfield  {author} {\bibinfo {author} {\bibfnamefont {J.}~\bibnamefont
  {Sun}}, \bibinfo {author} {\bibfnamefont {P.~S.}\ \bibnamefont {Carney}},\
  and\ \bibinfo {author} {\bibfnamefont {J.~C.}\ \bibnamefont {Schotland}},\
  }\bibfield  {title} {\bibinfo {title} {Strong tip effects in near-field
  scanning optical tomography},\ }\href@noop {} {\bibfield  {journal} {\bibinfo
   {journal} {Journal of Applied Physics}\ }\textbf {\bibinfo {volume} {102}}
  (\bibinfo {year} {2007})}\BibitemShut {NoStop}%
\bibitem [{\citenamefont {Wolf}(1969)}]{wolf1969three}%
  \BibitemOpen
  \bibfield  {author} {\bibinfo {author} {\bibfnamefont {E.}~\bibnamefont
  {Wolf}},\ }\bibfield  {title} {\bibinfo {title} {Three-dimensional structure
  determination of semi-transparent objects from holographic data},\
  }\href@noop {} {\bibfield  {journal} {\bibinfo  {journal} {Optics
  communications}\ }\textbf {\bibinfo {volume} {1}},\ \bibinfo {pages} {153}
  (\bibinfo {year} {1969})}\BibitemShut {NoStop}%
\bibitem [{\citenamefont {Weyl}(1919)}]{weyl1919ausbreitung}%
  \BibitemOpen
  \bibfield  {author} {\bibinfo {author} {\bibfnamefont {H.}~\bibnamefont
  {Weyl}},\ }\bibfield  {title} {\bibinfo {title} {Ausbreitung
  elektromagnetischer wellen {\"u}ber einem ebenen leiter},\ }\href@noop {}
  {\bibfield  {journal} {\bibinfo  {journal} {Annalen der Physik}\ }\textbf
  {\bibinfo {volume} {365}},\ \bibinfo {pages} {481} (\bibinfo {year}
  {1919})}\BibitemShut {NoStop}%
\bibitem [{cod()}]{code}%
  \BibitemOpen
  \bibfield  {title} {\bibinfo {title} {Code repository to support the
  finidings of this paper:}\ }\href {https://doi.org/10.5281/zenodo.17185502}
  {10.5281/zenodo.17185502}\BibitemShut {NoStop}%
\end{thebibliography}%

\end{document}